\documentclass[sigconf, nonacm]{acmart} %
\usepackage{balance}
\usepackage{textpos}
\usepackage[utf8]{inputenc}
\usepackage{amssymb}
\usepackage{bm}
\usepackage{graphicx}
\usepackage{amsfonts}
\usepackage{multicol}
\usepackage{color}
\usepackage{subcaption}
\usepackage{textcomp}

\makeatletter
\let\@authorsaddresses\@empty
\makeatother

\definecolor{Orange}{rgb}{0.9,0.5,0}
\definecolor{NavyBlue}{rgb}{0.1, 0.4, 0.8}
\definecolor{Magenta}{rgb}{0.8, 0.1, 0.6}
\definecolor{Green}{rgb}{0.1, 0.8, 0.3}
\definecolor{DarkGreen}{rgb}{0.0, 0.7, 0.2}
\definecolor{Brown}{rgb}{0.4, 0.3, 0.1}
\definecolor{Burgundy}{rgb}{0.5, 0.0, 0.13}
\definecolor{BrightCerulean}{rgb}{0.11, 0.67, 0.84}
\definecolor{BlueViolet}{rgb}{0.33,0.1,0.5}

\AtBeginDocument{%
  \providecommand\BibTeX{{%
    \normalfont B\kern-0.5em{\scshape i\kern-0.25em b}\kern-0.8em\TeX}}}

\setcopyright{acmcopyright}
\copyrightyear{2019}
\acmYear{2019}
\acmDOI{0}

\acmJournal{JACM}
\acmVolume{0}
\acmNumber{0}
\acmArticle{0}
\acmMonth{0}

\begin{document}

\title{Towards Characterizing and Limiting Information Exposure in DNN Layers}

\author{Fan Mo}
\affiliation{%
  \institution{Imperial College London}
  \streetaddress{}
  \city{}
  \country{}
}

\author{Ali Shahin Shamsabadi}
\affiliation{%
  \institution{Queen Mary University of London}
  \streetaddress{}
  \city{}
  \country{}
}

\author{Kleomenis Katevas}
\affiliation{%
  \institution{Imperial College London}
  \streetaddress{}
  \city{}
  \country{}
}

\author{Andrea Cavallaro}
\affiliation{%
  \institution{Queen Mary University of London}
  \streetaddress{}
  \city{}
  \country{}
}

\author{Hamed Haddadi}
\affiliation{%
  \institution{Imperial College London}
  \streetaddress{}
  \city{}
  \country{}
}

\begin{abstract}
  Pre-trained Deep Neural Network (DNN) models are increasingly used in smartphones and other user devices to enable prediction services, leading to potential disclosures of (sensitive) information from training data captured inside these models. Based on the concept of generalization error, we propose a framework to measure the amount of sensitive information memorized in each layer of a DNN. Our results show that, when considered individually, the last layers encode a larger amount of information from the training data compared to the first layers. We find that, while the neuron of convolutional layers can expose more (sensitive) information than that of fully connected layers, the same DNN architecture trained with different datasets has similar exposure per layer. We evaluate an architecture to protect the most sensitive layers within the memory limits of Trusted Execution Environment (TEE) against potential white-box membership inference attacks without the significant computational overhead.

\end{abstract}


\maketitle

\section{Introduction}

On-device DNNs have achieved impressive performance on a broad spectrum of services based on images, audio, and text. Examples include face recognition for authentication~\cite{vazquez2016face}, speech recognition for interaction~\cite{mcgraw2016personalized} and natural language processing for auto-correction~\cite{bellegarda2017unified}. However, DNNs memorize in their parameters information from the training data~\cite{zhang2016understanding, yeom2018privacy, zeiler2014visualizing}. Thus, keeping DNNs accessible in user devices leads to privacy concerns when training data contains sensitive information.

Previous works have shown that a reconstruction of the original input data is easier from the first layers of a DNN, when using for inference the layer's output (activation)~\cite{gu2018securing, osia2018private, osia2017hybrid}. In addition, the functionality of the parameters of each layer is different. For example, parameters of first layers trained (on images) output low-level features, whereas parameters of later layers learn higher level features, such as faces ~\cite{zeiler2014visualizing}.

We hypothesize that the memorization of sensitive information from training data differs across the layers of a DNN and, in this paper, present an approach to measure this sensitive information. We show that each layer behaves differently on the data they were trained on compared to data seen for the first time, by quantifying the generalization error (i.e. the expected distance between prediction accuracy of training data and test data~\cite{yeom2018privacy, shalev2010learnability}).
We further quantify the risk of sensitive information exposure of each layer as a function of its maximum and minimum possible generalization error. The larger the generalization error, the easier the inference of sensitive information from training set data. 

We perform experiments by training VGG-7~\cite{simonyan2014very} on three image datasets: MNIST~\cite{lecun2010mnist}, Fashion-MNIST~\cite{xiao2017fashion}, and CIFAR-10~\cite{krizhevsky2009learning}.
Our results show that last layers memorize more sensitive information about training data, and the risk of information exposure of a layer is independent of the dataset.

To protect the most sensitive layers from potential white-box attacks~\cite{melis2019exploiting, hitaj2017deep, nasr2018comprehensive}, we leverage a resource-limited Trusted Execution Environment (TEE)~\cite{chou2018faster, ohrimenko2016oblivious, hunt2018chiron} unit, Arm's TrustZone, as a protection example. Experiments are conducted by training last layers in the TEE and first layers outside the TEE. Results show that the overhead in memory, execution time and power consumption is minor, thus making it an affordable solution to protect a model from potential attacks.

\section{Proposed approach}

\subsection{Problem Definition}
\label{sec:def}

Let $M(\theta)$ be a DNN with $L$ layers, parameterized by $\theta=\{\mathbf{\theta}_{l}\}_{l=1}^L$, where $\mathbf{\theta}_{l}$ is the matrix with the parameters of layer $l$. Let $X=\{\mathbf{X}_{k}\}_{k=1}^{K}$ be the training set of $K$ images $\mathbf{X}_{k}$. Let $S=\{\mathbf{S}_{k}\}_{k=1}^{K_s} \subset X$, a randomly selected subset of $X$ with $K_s = K/2$, be the {\em private dataset} and $T=\{\mathbf{T}_{k}\}_{k=1}^{K_t} \subset X$, with $T \cap S = \emptyset$, be the non-private dataset.

As training $M(\theta)$ on $S$ might embed some information of $\mathbf{S}_{k}$ in the parameters of each layer, we aim to quantify the exposure of sensitive information in each $\mathbf{\theta}_{l}$. The sensitive information we are interested in analyzing is the absence or presence of any $\mathbf{S}_{k}$ in the training data.

\begin{figure*}
    \centering
    \includegraphics[width=2\columnwidth]{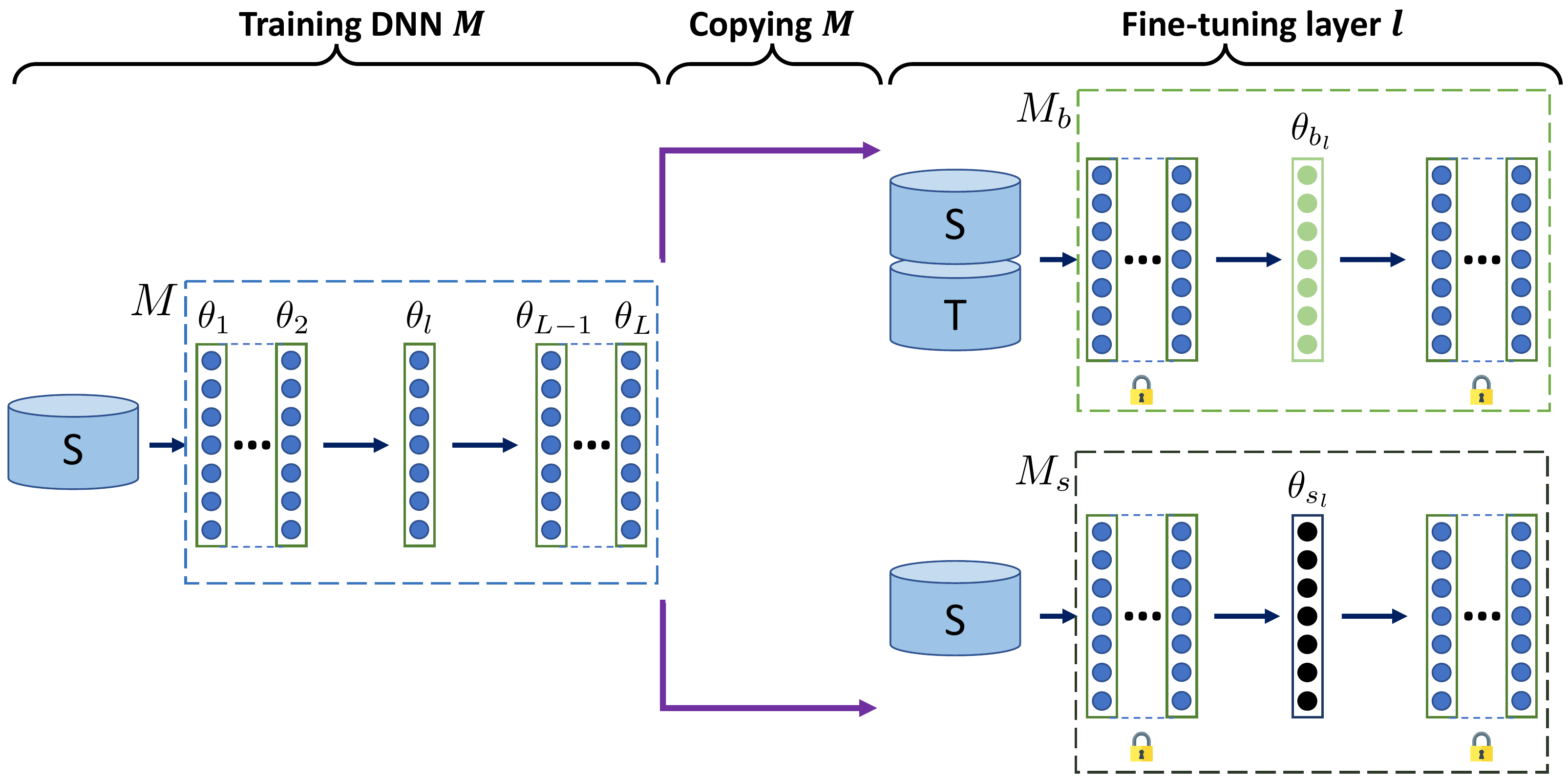} %
    \caption{The proposed framework for measuring the risk of exposing sensitive information  in a deep neural network $M$ trained on a private dataset $S$. $M_b$ and $M_s$ are obtained by fine-tuning the parameters of a target layer $l$ on the whole training set $X$ (i.e.~both $S$ and non-private training set $T$) and $S$, respectively.}
    \label{fig:privacy_metric_structure}
\end{figure*}
%

\subsection{Sensitive Information Exposure}

We leverage the fact that $M(\theta)$, trained on $S$, has a higher accuracy of predicting classes of data points from $S$ than from another dataset, $T$. The difference in prediction accuracy indicates the generalization error~\cite{yeom2018privacy, shalev2010learnability} of $M(\theta)$ and how easy is to recognize whether a data point ${X}_k$ was in $S$ during training. We define the risk of sensitive information exposure of each $\mathbf{\theta}_{l}$ based on the maximum and minimum possible generalization errors (see Figure~\ref{fig:privacy_metric_structure}). A larger difference in the maximum and minimum of generalization error could show the more sensitive information exposure which results in inferring more accurately the absence or presence of data in the training data (i.e. membership inference attack~\cite{yeom2018privacy}).

To obtain the maximum generalization error, we increase the chance of overfitting $\mathbf{\theta}_{l}$ to $S$ by fine-tuning $\mathbf{\theta}_{l}$ and by freezing parameters of other layers of $M$. We call this model $M_s(\theta_s)$. If $C (M_s, \mathbf{S}_{i})$ is the distance between $M_s$ and $\mathbf{S}_{i}$ measured by the cost function used in training, we quantify  $\epsilon_{s}$, the generalization error of $M_s(\theta_s)$, based on  its different behaviour on $S$ and $T$:
\begin{equation}
    \label{eq:genS}
    \epsilon_{s}=\mathbb{E}_{ \mathbf{T}_{i} \in T} [C (M_s, \mathbf{T}_{i})] - \mathbb{E}_{\mathbf{S}_{i} \in S} [C (M_s, \mathbf{S}_{i})],
\end{equation}
where $\mathbb{E}[\cdot]$ is the mathematical expectation.

To obtain the minimum generalization error without forgetting $S$, we create a baseline $M_b(\theta_b)$ by fine-tuning $\mathbf{\theta}_{l}$ on $X$ and by freezing the parameters of the other layers of $M$. This fine-tuning makes $\mathbf{\theta}_{b_l}$ generalized on both $T$ and $S$, which can be quantified as:
\begin{equation}
\label{eq:genb}
    \epsilon_{b}=\mathbb{E}_{ \mathbf{T}_{i} \in T} [C (M_b, \mathbf{T}_{i})] - \mathbb{E}_{\mathbf{S}_{i} \in S} [C (M_b, \mathbf{S}_{i})].
\end{equation}

$M(\theta)$, $M_s(\theta_s)$ and $M_b(\theta_b)$ share the same layers, except the target layer ${l}$. Therefore, the differences in each pair of %
\begin{equation*}
    \{\mathbb{E}_{ \mathbf{T}_{i} \in T} [C (M_s, \mathbf{T}_{i})], \mathbb{E}_{ \mathbf{T}_{i} \in T} [C (M_b, \mathbf{T}_{i})]\},
\end{equation*}
and
\begin{equation*}
    \{\mathbb{E}_{ \mathbf{S}_{i} \in S} [C (M_s, \mathbf{S}_{i})], \mathbb{E}_{ \mathbf{S}_{i} \in S} [C (M_b, \mathbf{S}_{i})]\},
\end{equation*}
are due to different parameters of layer $l$.

We therefore quantify $\mathcal{R}^{M_s}$, the risk of sensitive information exposure of layer ${l}$,  by comparing the generalization error of $M_s$ and $M_b$: 
\begin{equation}
\label{eq:Leak}
    \mathcal{R}^{M_s}=\frac{\epsilon_{s}-\epsilon_{b}}{\epsilon_{s}}.
\end{equation}

The larger $\mathcal{R}^{M_s}$, the higher the risk of exposing sensitive information.

\section{Measuring information exposure}

\subsection{Model and Datasets}

We use VGG-7 as the DNN $M$, which has six convolutional layers followed by one fully connected layer (16C3-16C3-MP-32C3-32C3-MP-32C3-32C3-MP-64FC-10SM). Each layer is followed by Rectifier Linear Unit (ReLU)~\cite{nair2010rectified} activation function.

We use three datasets: MNIST, Fashion-MNIST, and CIFAR-10. MNIST includes 60k training images of $28 \times 28 \times 1$ handwritten digits of 10 classes (i.e. 0 to 9). Fashion-MNIST contains 60k $28 \times 28 \times 1$ images of 10 classes of clothing, namely T-shirt/top, trouser, pullover, dress, coat, sandal, shirt, sneaker, bag, and ankle boot. CIFAR-10 includes 50k training $32 \times 32 \times 3$ images of 10 classes including airplane, automobile, bird, cat, deer, dog, frog, horse, ship, and truck.

We split each training set into set $S$ and set $T$, as explained in Sec.~\ref{sec:def}. We use 20 epochs for MNIST, 40 epochs for Fashion-MNIST, and 60 epochs for CIFAR-10. The accuracy of VGG-7 on MNIST, Fashion-MNIST, and CIFAR-10 is 99.29\%, 90.55\%, and 71.63\%, respectively. We then fine-tune $M$ as $M_s$ and $M_b$ with 10 epochs for MNIST, 20 epochs for Fashion-MNIST, and 30 epochs for CIFAR-10.

\begin{figure*}
\centering
   \begin{subfigure}{0.6\columnwidth} 
     \includegraphics[scale=0.6]{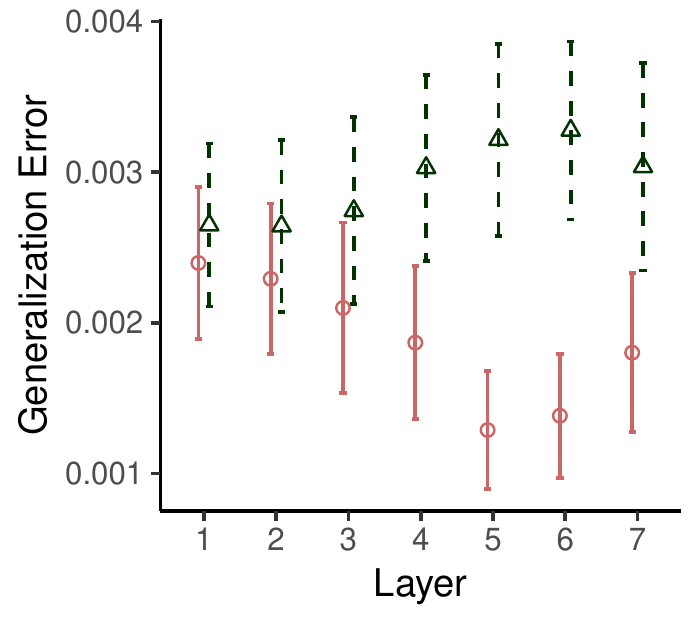}
     \caption{MNIST}\label{fig:mnist_gen}
   \end{subfigure}
   \begin{subfigure}{0.6\columnwidth} 
     \includegraphics[scale=0.6]{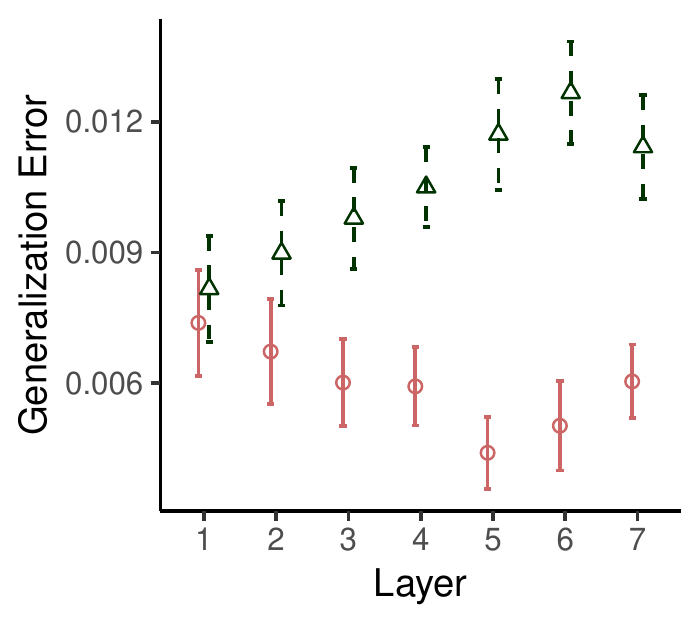}
     \caption{Fashion-MNIST}\label{fig:fashionmnist_gen}
   \end{subfigure}
    \begin{subfigure}{0.72\columnwidth} \centering
     \includegraphics[scale=0.6]{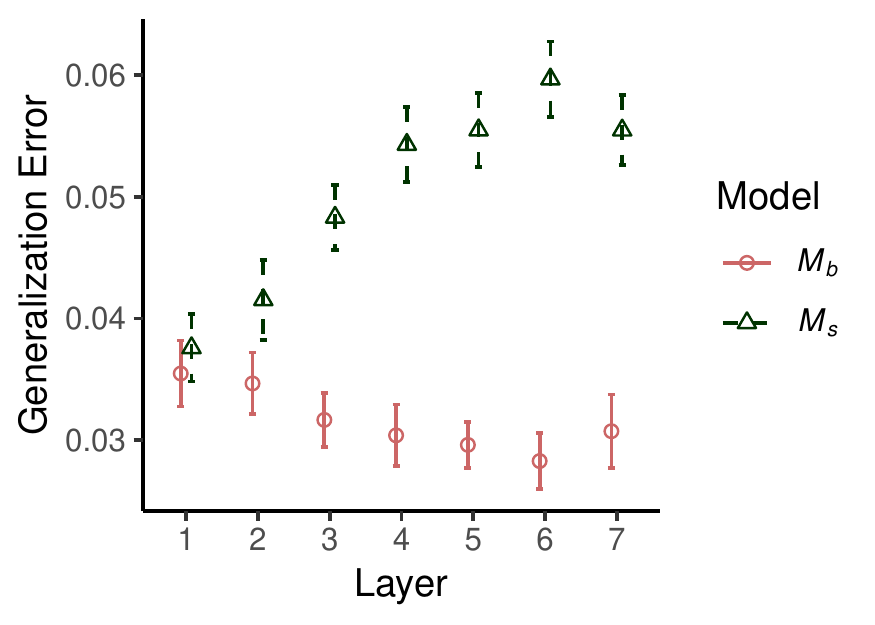}
     \caption{CIFAR-10}\label{fig:cifar_gen}
   \end{subfigure}
    \caption{Generalization errors of $M_s$ and $M_b$ trained on half of the training set, $S$, of (a) MNIST, (b) Fashion-MNIST  and (c)~CIFAR-10 for fine-tuning each target layer. Error bars represent 95\% confidence intervals.}
     \label{fig:model_generalization}
\end{figure*}

\subsection{Results and Discussion}

\textbf{Generalization error}. Figure~\ref{fig:model_generalization} shows the generalization errors of $M_s$ and $M_b$. For all three datasets, the baseline model $M_b$, as expected, has higher generalization errors than the model $M_s$, whose layer $l$ is overfitted to dataset $S$, while the generalization error of CIFAR-10 is greater than that of Fashion-MNIST that in turn is greater than that of MNIST. A more complex dataset (e.g.~CIFAR-10) is associated to a larger difference between $S$ and $T$ compared to a less complex dataset (e.g.~MNIST), so it is harder to generalize the model to predict $T$ by training with $S$.

As we go through the convolutional layers, the generalization error of $M_s$ increases, while the generalization error of $M_b$ decreases until the $5^{th}$ or $6^{th}$ layer. A possible explanation is that first layers memorize generic information (e.g.~colors, and corners), whereas last layers memorize more specific information that can be used to identify a specific image. For example, fine-tuning the last layers using $S$ leads $M_s$ to memorize specific information of $S$, which consequentially increases the generalization errors of $M_s$ when predicting $T$.

\begin{figure}
    \centering
     \includegraphics[width=1\columnwidth]{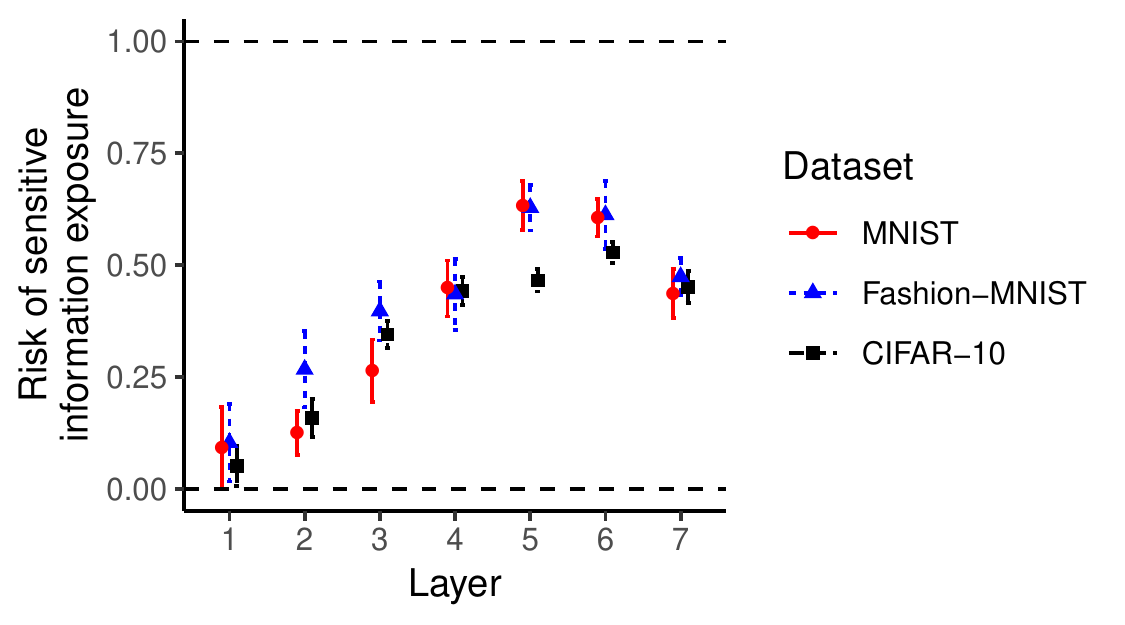} %
     \caption{The risk of sensitive information exposure of VGG-7 per layer on MNIST, Fashion-MNIST and CIFAR-10. Error bars represent 95\% confidence intervals.}
     \label{fig:privacy_by_layers}
\end{figure}

\noindent 
\textbf{Sensitive information exposure.} Figure \ref{fig:privacy_by_layers} shows the risk of sensitive information exposure for each layer of VGG-7 on all three datasets. The first layer has the lowest risk, and the risk increases as we go through the layers, with the last convolutional layer having the highest sensitive information exposure, which is 0.63 for both MNIST and Fashion-MNIST and 0.5 for CIFAR-10. This confirms the bigger derivation of the generalization error of $M_s$ from $M_b$ in the last layers than the first layers. In addition, the order of layers in terms of sensitive information exposure is almost the same across all three datasets.

\begin{figure}
    \centering
     \includegraphics[width=1\columnwidth]{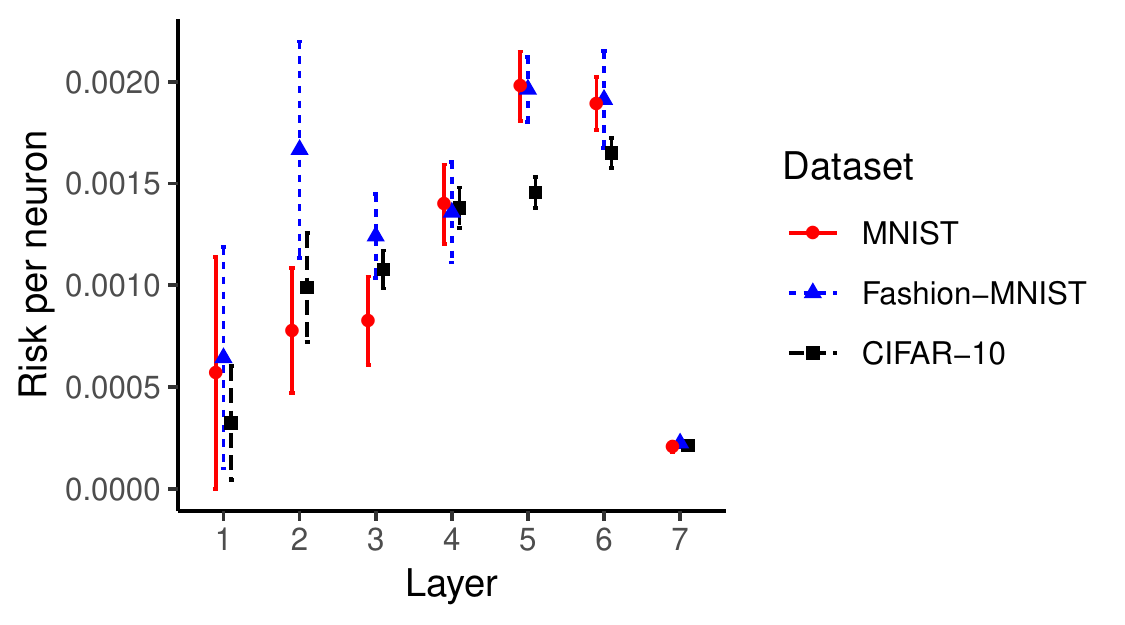} %
     \caption{Risk per neuron for each layer on MNIST, Fashion-MNIST and CIFAR-10. Error bars represent 95\% confidence intervals.}
     \label{fig:privacy_per_neuron}
\end{figure}

We also compute the risk per neuron for each layer by normalizing the risk of sensitive information exposure by the total number of neurons of the layer (Figure~\ref{fig:privacy_per_neuron}). The results show the risk per neuron increases as we move through convolutional layers. Neurons in the late convolutional layers have high capabilities in memorizing sensitive information, whereas the fully connected layer (layer $7$) has a much smaller risk per neuron.

\section{Trusted Execution Environment}

\subsection{Setup}

\begin{figure}[t]
    \centering
    \includegraphics[width=0.7\columnwidth]{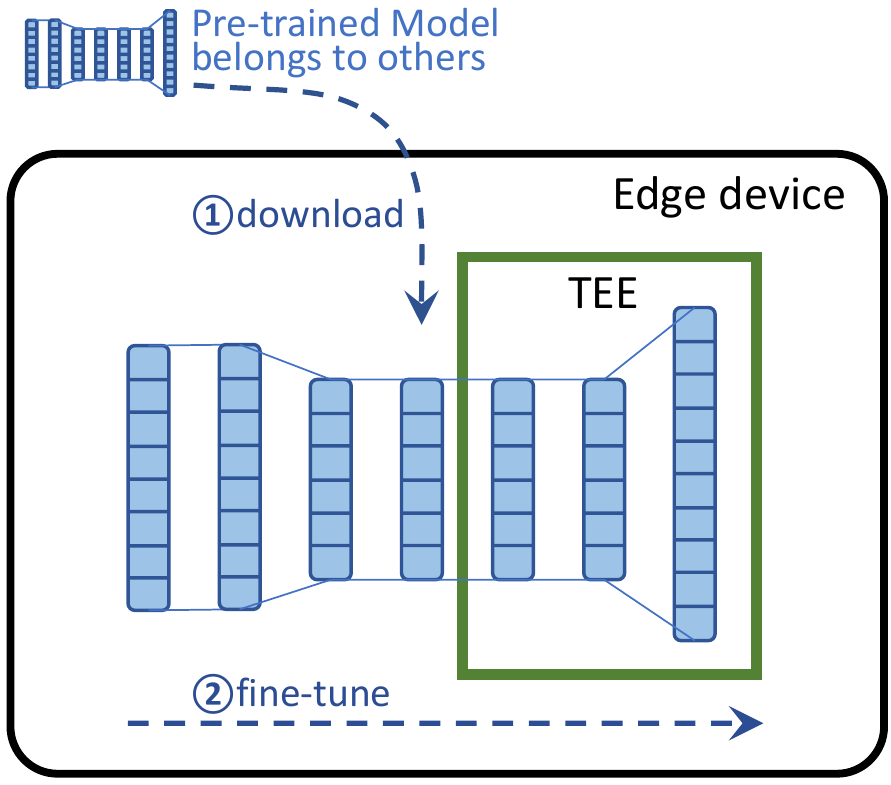}
    \caption{Using a TEE to protect the most sensitive layers (last layers) of an on-device deep neural network.}
    \label{fig:scenario}
\end{figure}

\begin{figure}
    \centering
    \begin{subfigure}{0.5\columnwidth} 
     \includegraphics[scale=0.72]{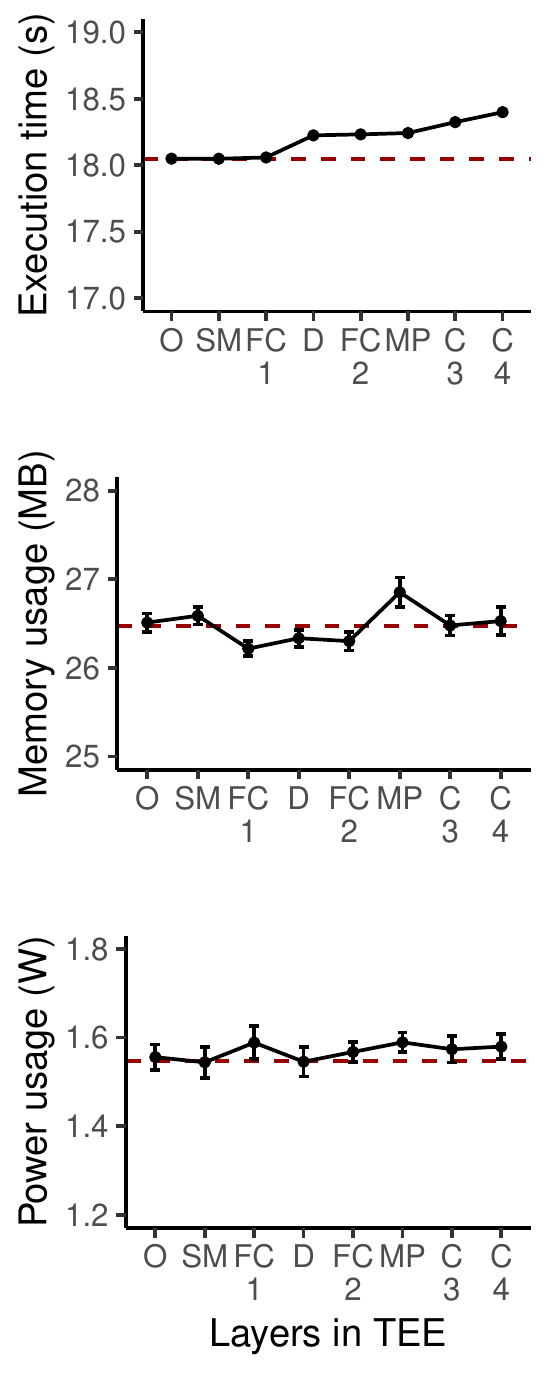}
     \caption{MNIST}\label{fig:tee_mnist}
   \end{subfigure}%
    \begin{subfigure}{0.5\columnwidth}
     \includegraphics[scale=0.72]{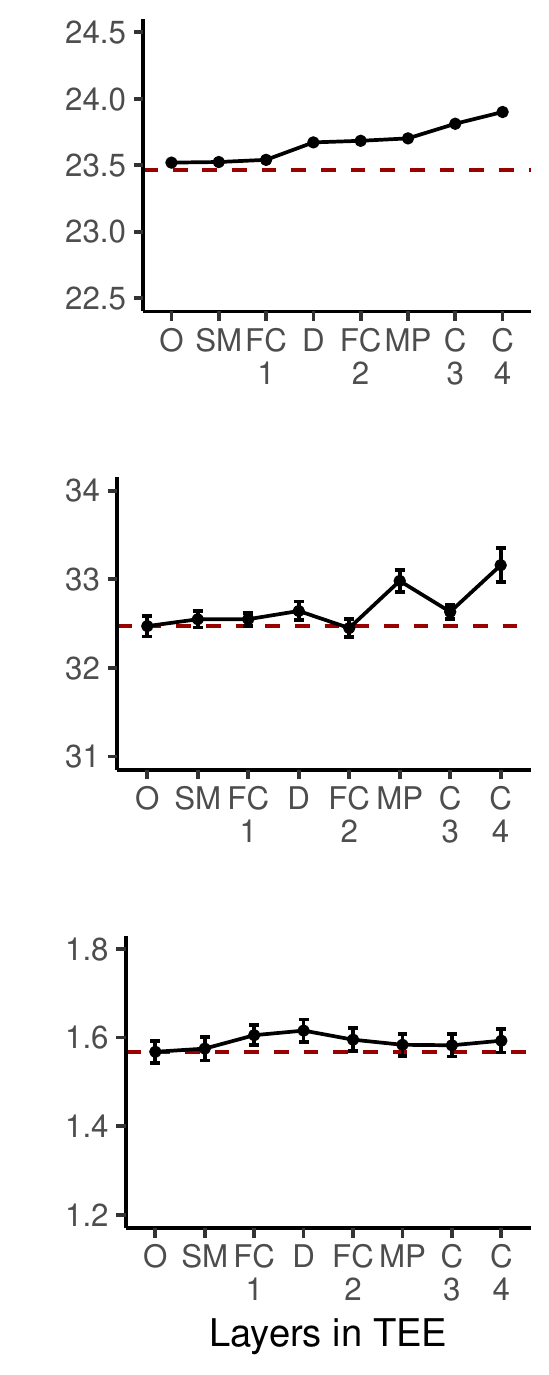}
     \caption{CIFAR-10}\label{fig:tee_cifar}
   \end{subfigure}%
    \caption{Execution time, memory usage and power usage for protecting layers of VGG-7 trained on MNIST (left column) and CIFAR-10 dataset (right column) using the TrustZone of device. The x-axis corresponds to several last layers included in the TrustZone. \emph{O} refers to the calculation of cost function; \emph{SM}, \emph{FC}, \emph{D}, \emph{MP}, and \emph{C} refer to the softmax, fully connected, dropout, maxpooling, convolutional layers of VGG-7. Number of layers with trainable parameters in the TrustZone are 1, 2, 3, and 4. The dash line represent the baseline, which runs all the layers outside the TrustZone. Error bars represent 95\% confidence intervals.}
    \label{fig:tee_cost_together}
\end{figure}

In this section, we develop an implementation and evaluate the cost of protecting the last layers of an on-device DNN during fine-tuning by deploying them in the TrustZone of a device (see Figure~\ref{fig:scenario}).
TrustZone is ARM's TEE implementation that establishes a private region on the main processor. Both hardware and software approaches isolate this region to allow trusted execution. As TEEs are usually small, we only protect the most sensitive layers of the model and use the normal execution environment for the other layers.

We use Darknet~\cite{darknet13} DNN library in Open Portable TEE (OP-TEE)\footnote{https://www.op-tee.org}, a TEE framework based on TrustZone, of a Raspberry Pi 3 Model B. This model of Raspberry Pi 3 runs instances of OP-TEE with 16 mebibytes (MiB) TEE's memory. The choice of Darknet~\cite{darknet13} is due to its high performance and small dependencies. The scripts we used in our evaluation are available online\footnote{https://github.com/mofanv/darknetp}.

We fine-tune the pre-trained VGG-7 (from the previous section) with MNIST and CIFAR-10, respectively.
Continuous layers are deployed in the TrustZone from the end for simplicity, including both layers with (i.e.~the convolutional and fully connected layer) and without (i.e.~the dropout and maxpooling layer) trainable parameters.

\subsection{Results and Discussion}

Figure~\ref{fig:tee_cost_together} shows the execution time (in seconds), memory usage (in MB), and power consumption (in Watt, using RuiDeng USB Tester (UM25C)\footnote{http://www.ruidengkeji.com}) of securing a part of the DNN in the TrustZone, starting from the last layer, and continuing adding layers until the maximum number of layers the zone can hold. 

The resulting execution times are MNIST: $F_{(7,232)}=3658$, $p<0.001$; CIFAR-10: $F_{(7,232)}=2396$, $p<0.001$ and memory usage is MNIST: $F_{(7,232)}=11.62$, $p<0.001$; CIFAR-10: $F_{(7,232)}=20.01$, $p<0.001$. The increase however is small compared to the baseline (Execution time: 1.94\% for MNIST and 1.62\% for CIFAR-10; Memory usage: 2.43\% for MNIST and 2.19\% for CIFAR-10). Moreover, running layers in the TrustZone did not significantly influence the power usage (MNIST: $F_{(7,232)}=1.49$, $p=0.170$; CIFAR-10: $F_{(7,232)}=1.61$, $p=0.132$).

Specifically, deploying the dropout layer and the maxpooling layer in the TEE increases both the execution time and memory usage. The reason is that these two types of layers have no trainable parameters, and for Darknet, the dropout and maxpooling are directly operated based on trainable parameters of their front layer. Therefore, to run these two types of layers in the TEE, their front layer (i.e.~fully connected/convolutional layers) needs to be copied into the TEE, which increases the cost. For layers with parameters that we aim to protect (1, 2, 3, and 4 in Figure~\ref{fig:tee_cost_together}), deploying fully connected layers (i.e. 1, 2) in the TEE does not increase the execution time accumulated on first layers, and does not increase the memory usage. Deploying convolutional layers (i.e. 3 and 4) leads to an increase of execution time but does not increase memory usage when using MNIST. The second convolutional layer (i.e. 4) only increases memory usage when using CIFAR-10. However, exhausting the most available memory of the TEE can also cause an increase of overhead, so the reason for this increment of memory usage needs more analysis. Overall, for our implementation, protecting fully connected and convolutional layers has lower costs than other layers without trainable parameters with the TEE.

\section{Conclusion}

We proposed a method to measure the exposure of sensitive information in each layer of a pre-trained DNN model. We showed that the closer the layer is to the output, the higher the likelihood that sensitive information of training data is exposed, which is opposite to the exposure risk of layers' activation from test data~\cite{gu2018securing}. 
We evaluated the use of TEE to protect individual sensitive layers (i.e.~the last layers) of a deployed DNN. The results show that TEE has a promising performance at low cost.

Future work includes investigating the advantages of protecting the later layers of a DNN against, among others, white-box membership inference attacks~\cite{nasr2018comprehensive}.

\section*{Acknowledgement}
We acknowledge the constructive advice and feedback from Soteris Demetriou and Ilias Leontiadis. The research in this paper is supported by grants from the EPSRC (Databox EP/N028260/1, DADA EP/R03351X/1, and HDI EP/R045178/1).

\bibliographystyle{ACM-Reference-Format}
\balance
\small
\bibliography{references}

\end{document}